\documentclass[aps,prd,twocolumn,superscriptaddress,showpacs,showkeys,
nofootinbib]{revtex4-1}

\usepackage{latexsym}
\usepackage{amsmath}
\usepackage{amssymb}
\usepackage{amsfonts}

\usepackage{color}
\usepackage{soul}

\usepackage{supertabular} 
\usepackage{placeins}
\usepackage{epsfig}
\usepackage{graphicx}

\def\tstrut{\vrule height2.25ex depth0pt width0pt} 

\begin{document}


\title{Threshold effects in P-wave bottom-strange mesons}



\author{Pablo G. Ortega}
\email[]{pgortega@ific.uv.es}
\affiliation{Instituto de F\'isica Corpuscular (IFIC), Centro Mixto 
CSIC-Universidad de Valencia, ES-46071 Valencia, Spain}

\author{Jorge Segovia}
\email[]{jorge.segovia@tum.de}
\affiliation{Physik-Department, Technische Universit\"at M\"unchen, 
James-Franck-Str.~1, 85748 Garching, Germany}

\author{David R. Entem}
\email[]{entem@usal.es}
\author{Francisco Fern\'andez}
\email[]{fdz@usal.es}
\affiliation{Grupo de F\'isica Nuclear and Instituto Universitario de 
F\'isica Fundamental y Matem\'aticas (IUFFyM), Universidad de Salamanca, 
E-37008 Salamanca, Spain}


\date{\today}

\begin{abstract}
Using a nonrelativistic constituent quark model in which the degrees of freedom 
are quark-antiquark and meson-meson components, we have recently shown that the 
$D^{(\ast)}K$ thresholds play an important role in lowering the mass of the 
physical $D_{s0}^{\ast}(2317)$ and $D_{s1}(2460)$ states. This observation is 
also supported by other theoretical approaches such as lattice-regularised QCD 
or chiral unitary theory in coupled channels. Herein, we extend our computation 
to the lowest $P$-wave $B_{s}$ mesons, taking into account the corresponding 
$J^{P} = 0^{+}$, $1^{+}$ and $2^{+}$ bottom-strange states predicted by the 
naive quark model and the $BK$ and $B^{\ast}K$ thresholds. We assume that 
mixing with $B_{s}^{(\ast)}\eta$ and isospin-violating decays to 
$B_{s}^{(\ast)}\pi$ are negligible. This computation is important because there 
is no experimental data in the $b\bar{s}$ sector for the equivalent 
$j_{q}^{P}=1/2^{+}$ ($D_{s0}^{\ast}(2317)$, $D_{s1}(2460)$) heavy-quark 
multiplet and, as it has been seen in the $c\bar s$ sector, the naive 
theoretical result can be wrong by more than $100\,{\rm MeV}$. Our calculation 
allows to introduce the coupling with the $D$-wave $B^{\ast}K$ channel and 
to compute the probabilities associated with the different Fock components 
of the physical state.
\end{abstract}



\maketitle



\section{Introduction}
\label{sec:introduction}

Despite many of the $B_{s}$ states should be accessible by the $B$-factories 
(CLEO, BaBar, Belle) and also by the proton--anti-proton colliders (CDF and 
D0), much of the $b\bar s$ excitation spectrum remains to be observed. Only the 
ground $S$-wave spin-singlet and spin-triplet states ($B_{s}$ and 
$B_{s}^{\ast}$) and the orbitally excited $B_{s1}(5830)$ 
and $B_{s2}^{\ast}(5840)$ mesons are presently well 
established~\cite{Agashe:2014kda}. It is expected that the situation will 
change in the near future thanks to the LHCb experiment and to the future 
high-luminosity flavour and $p-\bar p$ facilities. 
 
In the heavy quark limit ($m_{Q}\to\infty$), flavour and spin symmetries hold 
and the open-flavoured heavy mesons can be classified in doublets. This is 
because in the heavy quark limit the spin $s_{Q}$ of the heavy quark and the 
total angular momentum of the light quark $j_{q}$ decouple and are separately 
conserved in strong interaction processes~\cite{Isgur:1991wq}. Therefore, $Q\bar 
q$ mesons can be classified according to the value of $j_{q}$, and can be 
collected in doublets; the two states of each doublet are spin partners with 
total spin $J = j_{q} \pm \frac{1}{2}$ and parity $P = (-1)^{\ell+1}$, with 
$\ell$ the orbital angular momentum of the light degrees of freedom such as 
$\vec{j}_q = \vec{\ell} + \vec{s}_q$ ($s_{q}$ is the light antiquark spin).

The well established $B_{s1}(5830)$ and $B_{s2}^{\ast}(5840)$ states belong to 
the $j_{q}^{P}=3/2^{+}$ doublet and therefore the unambiguous experimental 
determination of the $b\bar s$ $j_{q}^{P}=1/2^{+}$ doublet is of particular 
interest because, following the predictions of heavy quark symmetry(HQS), they 
should be almost degenerated and broad. However, we know that the experimental 
values of the masses and widths of the 
$D_{s0}^{\ast}(2317)$~\cite{Aubert:2003fg} and 
$D_{s1}(2460)$~\cite{Besson:2003cp} mesons, which belong to the same multiplet 
but in the $c\bar s$ sector, do not accommodate into the theoretical estimates.

We have recently argued in Ref.~\cite{Ortega:2016mms} that the disagreement 
between theory and experiment for the $D_{s0}^{\ast}(2317)$ and $D_{s1}(2460)$ 
mesons can be explained as a delicate compromise between the bare masses  and 
the coupling with their respective $DK$ and $D^*K$ thresholds. The study of 
Ref.~\cite{Ortega:2016mms} taught us that the traditional quark model do not 
reproduce the masses and widths of the $D_{s0}^{\ast}(2317)$ and $D_{s1}(2460)$ 
mesons. When, following the proposal of Ref.~\cite{Lakhina:2006fy}, we include 
one-loop corrections to the one-gluon exchange
potential as derived by Gupta {\it et al.}~\cite{Gupta:1981pd},  these $\alpha_s^2$
corrections affect basically the $0^+$ $D_{s0}^{\ast}(2317)$, bringing this state closer to the $DK$ threshold.
However, the same  correction leaves the $D_{s1}(2460)$ almost degenerated with the $D_{s1}(2536)$. 
It is the coupling with the $DK$ and 
$D^{\ast}K$ channels which lowers the bare masses of the $D_{s0}^{\ast}(2317)$ 
and $D_{s1}(2460)$ mesons to the observed experimental values. Since their 
masses are below their respective $DK$ and $D^{\ast}K$ thresholds, their 
open-flavoured strong decays are forbidden and thus the states are narrower than 
expected theoretically. Similar conclusions were drawn by dynamical 
coupled-channel approaches~\cite{vanBeveren:2003kd, 
vanBeveren:2003jv} and by lattice-regularised QCD 
computations~\cite{Mohler:2013rwa, Lang:2014zaa}.

In this work we closely follow  Ref.~\cite{Ortega:2016mms} and address the 
mass-shifts, due to the $B^{(\ast)}K$ thresholds, of the lowest lying $P$-wave 
$b\bar s$ states with total spin and parity quantum numbers $J^P = 0^+$, $1^+$, 
$2^+$.

A first lattice-regularised QCD study of the lowest-lying $P$-wave 
bottom-strange mesons in which $B^{(\ast)}K$ operators are explicitly 
incorporated in the interpolator basis has been recently 
released~\cite{Lang:2015hza}. We shall mainly compare our results with the ones 
obtained by them, but a number of phenomenological model and effective field 
theory (EFT) mass determinations of the same states can be found in 
Tables~\ref{tab:Bs0Ene} and~\ref{tab:Bs1Ene}.

\begin{table}[!t]
\begin{center}
\begin{tabular}{lcl}
\hline
\hline
\multicolumn{3}{c}{$B_{s0}^{\ast}$ meson} \\
\hline
Relativistic quark model & \cite{DiPierro:2001dwf} & $5804$ \\
Relativistic quark model & \cite{Ebert:2009ua}     & $5833$ \\
Relativistic quark model & \cite{Sun:2014wea}      & $5830$ \\
Relativized quark model  & \cite{Godfrey:2016nwn}  & $5805$ \\
\hline
Bardeen, Eichten, Hill  & \cite{Bardeen:2003kt}    & $5718(35)$ \\
Q.-F. L\"u {\it et al.} & \cite{Lu:2016bbk}        & $5756$ \\
\hline
LQCD: $q\bar{q}$ + $BK$ & \cite{Lang:2015hza}   & $5713(11)(19)$ \\
LQCD: $q\bar{q}$        & \cite{Gregory:2010gm} & $5752(16)(5)(25)$ \\
\hline
Covariant (U)ChPT & \cite{Altenbuchinger:2013vwa} & $5726(28)$ \\
NLO UHMChPT       & \cite{Cleven:2010aw}          & $5696(20)(30)$ \\
LO UChPT          & \cite{Guo:2006fu,Guo:2006rp}  & $5725(39)$ \\
LO $\chi$-SU(3)   & \cite{Kolomeitsev:2003ac}     & $5643$ \\
HQET + ChPT       & \cite{Colangelo:2012xi}       & $5706.6(1.2)$ \\
\hline
\hline
\end{tabular}
\caption{\label{tab:Bs0Ene} Mass, in MeV, of the $B_{s0}^{\ast}$ meson 
predicted by different theoretical approaches.}
\end{center}
\end{table}

\begin{table}[!t]
\begin{center}
\scalebox{0.92}{\begin{tabular}{lcll}
\hline
\hline
\multicolumn{3}{c}{$B_{s1}^{\prime}$ meson} \\
\hline
Relativistic quark model & \cite{DiPierro:2001dwf} & $5842$ \\
Relativistic quark model & \cite{Ebert:2009ua}     & $5865$ \\
Relativistic quark model & \cite{Sun:2014wea}      & $5858$ \\
Relativized quark model  & \cite{Godfrey:2016nwn}  & $5822$ \\
\hline
Bardeen, Eichten, Hill  & \cite{Bardeen:2003kt}    & $5765(35)$ \\
Q.-F. L\"u {\it et al.} & \cite{Lu:2016bbk}        & $5801$ \\
\hline
LQCD: $q\bar{q}$ + $B^{\ast}K(S)$ & \cite{Lang:2015hza}   & $5750(17)(19)$ \\
LQCD: $q\bar{q}$        & \cite{Gregory:2010gm} & $5806(15)(5)(25)$ \\
\hline
Covariant (U)ChPT & \cite{Altenbuchinger:2013vwa} & $5778(26)$ \\
NLO UHMChPT       & \cite{Cleven:2010aw}          & $5742(20)(30)$ \\
LO UChPT          & \cite{Guo:2006fu,Guo:2006rp}  & $5778(7)$ \\
LO $\chi$-SU(3)   & \cite{Kolomeitsev:2003ac}     & $5690$ \\
HQET + ChPT       & \cite{Colangelo:2012xi}       & $5765.6(1.2)$ \\
\hline
\hline
\end{tabular}}
\caption{\label{tab:Bs1Ene} Mass, in MeV, of the $B_{s1}^{\prime}$ meson 
predicted by different theoretical approaches.}
\end{center}
\end{table}

The first conclusion one can obtain from such tables is that the naive quark 
models, including the recent calculation of Godfrey {\it et al.}, predict 
masses above the $BK$ and $B^{\ast}K$ thresholds, respectively. The authors of 
Ref.~\cite{Lu:2016bbk} use a nonrelativistic quark model which include 
$\alpha_{s}^{2}$-corrections. Taken a set of parameters fitted to the charm 
sector, they succeed to bring the $0^{+}$ $b\bar{s}$ state below the $BK$ 
threshold. To break the degeneration between the $B_{s1}^{\prime}$ and $B_{s1}$ 
mesons, they adjust a mixing angle that also place the $B_{s1}^{\prime}$ below 
the $B^{\ast}K$ threshold. The result of Bardeen {\it et al.} is calculated 
using a set of parameters fitted to the $D_{s0}^{\ast}(2317)$. 

The second conclusion one can make is that the results of lattice computations, 
which take explicitly $B^{(\ast)}K$ operators in the interpolator basis, and 
the estimates of EFTs, that study dynamically generated bound states from the 
$B^{(\ast)}K$ scattering, are $80\,{\rm MeV}$ lower than the quark model 
predictions.

Despite the significant progress made by lattice calculations incorporating 
open-flavoured thresholds, two main drawbacks remain: i) the thresholds are 
added only as $S$-wave channels and ii) no statement can be made about 
the probabilities of the different Fock components in the physical state. Our 
approach solves these two issues and allows e.g. to introduce the coupling with 
the $D$-wave $B^{\ast}K$ channel in the $1^{+}$ $b\bar{s}$ sector and to 
compute the amount of $B^{(\ast)}K$ component in the meson.

Our theoretical framework is a nonrelativistic constituent quark model in 
which quark-antiquark and meson-meson degrees of freedom are incorporated. The 
constituent quark model (CQM) was proposed in Ref.~\cite{Vijande:2004he} (see 
references~\cite{Valcarce:2005em} and~\cite{Segovia:2013wma} for reviews).

In order to keep the predictive power of the formalism we do not change any 
model parameter. Moreover, it is worth to emphasize here that the quark model 
has been applied to a wide range of hadronic observables, e.g. 
Refs.~\cite{Segovia:2009zz, Segovia:2011zza, Segovia:2013kg, Segovia:2013sxa, 
Segovia:2014mca, Segovia:2015dia, Ortega:2010qq, Ortega:2012rs}, and thus the 
model parameters are completely constrained.

The manuscript is arranged as follows. In Sec.~\ref{sec:theory} we describe 
briefly the main properties of our theoretical formalism. 
Section~\ref{sec:Bs} is devoted to present our results for the lowest lying 
$P$-wave $B_{s}$ states with total spin and parity quantum numbers $J^P = 
0^+$, $1^+$, $2^+$. We finish summarizing and giving some conclusions in 
Sec.~\ref{sec:epilogue}.


\section{Constituent quark model}
\label{sec:theory}

Constituent light quark masses and Goldstone-boson exchanges, which are 
consequences of dynamical chiral symmetry breaking in Quantum Chromodynamics 
(QCD), together with the perturbative one-gluon exchange and a nonperturbative 
confining interaction are the main pieces of our constituent quark 
model~\cite{Vijande:2004he, Segovia:2013wma}.

A simple Lagrangian invariant under chiral transformations can be 
written in the following form~\cite{Diakonov:2002fq}
\begin{equation}
{\mathcal L} = \bar{\psi}(i\, {\slash\!\!\! \partial} 
-M(q^{2})U^{\gamma_{5}})\,\psi  \,,
\end{equation}
where $M(q^2)$ is the dynamical (constituent) quark mass and $U^{\gamma_5} = 
e^{i\lambda _{a}\phi ^{a}\gamma _{5}/f_{\pi}}$ is the matrix of Goldstone-boson 
fields that can be expanded as
\begin{equation}
U^{\gamma _{5}} = 1 + \frac{i}{f_{\pi}} \gamma^{5} \lambda^{a} \pi^{a} - 
\frac{1}{2f_{\pi}^{2}} \pi^{a} \pi^{a} + \ldots
\end{equation}
The first term of the expansion generates the constituent quark mass while the
second gives rise to a one-boson exchange interaction between quarks. The
main contribution of the third term comes from the two-pion exchange which
has been simulated by means of a scalar-meson exchange potential.

In the heavy quark sector chiral symmetry is explicitly broken and 
Goldstone-boson exchanges do not appear. However, it constrains the model 
parameters through the light-meson phenomenology~\cite{Segovia:2008zza} and 
provides a natural way to incorporate the pion exchange interaction in the 
molecular dynamics.

The one-gluon exchange (OGE) potential is generated from the vertex Lagrangian
\begin{equation}
{\mathcal L}_{qqg} = i\sqrt{4\pi\alpha_{s}} \,\, \bar{\psi}\, 
\gamma_{\mu}\, G^{\mu}_{c}\, \lambda^{c}\, \psi \,,
\label{Lqqg}
\end{equation}
where $\lambda^{c}$ are the $SU(3)$ colour matrices and $G^{\mu}_{c}$ is the
gluon field. The resulting potential contains central, tensor and spin-orbit
contributions.

To improve the description of the open-flavour mesons, we follow the proposal 
of Ref.~\cite{Lakhina:2006fy} and include one-loop corrections to the OGE 
potential as derived by Gupta {\it et al.}~\cite{Gupta:1981pd}. These 
corrections show a spin-dependent term which affects only mesons with different 
flavour quarks.

It is well known that multi-gluon exchanges produce an attractive linearly 
rising potential proportional to the distance between infinite-heavy quarks. 
However, sea quarks are also important ingredients of the strong interaction 
dynamics that contribute to the screening of the rising potential at low 
momenta and eventually to the breaking of the quark-antiquark binding 
string~\cite{Bali:2005fu}. Our model tries to mimic this behaviour using the 
following expression:
\begin{equation}
V_{\rm CON}(\vec{r}\,)=\left[-a_{c}(1-e^{-\mu_{c}r})+\Delta \right] 
(\vec{\lambda}_{q}^{c}\cdot\vec{\lambda}_{\bar{q}}^{c}) \,,
\label{eq:conf}
\end{equation}
where $a_{c}$ and $\mu_{c}$ are model parameters. 

Explicit expressions for all the potentials and the value of the model 
parameters can be found in Ref.~\cite{Vijande:2004he}, updated
in Ref.~\cite{Segovia:2008zz}. Meson eigenenergies and 
eigenstates are obtained by solving the Schr\"odinger equation using the 
Gaussian Expansion Method~\cite{Hiyama:2003cu}. 

The quark-antiquark bound state can be strongly influenced by nearby multiquark 
channels. In this work, we follow Refs.~\cite{Ortega:2010qq, Ortega:2012rs} to 
study this effect in the spectrum of the bottom-strange mesons and thus we need 
to assume that the hadronic state is given by
\begin{equation} 
| \Psi \rangle = \sum_\alpha c_\alpha | \psi_\alpha \rangle
+ \sum_\beta \chi_\beta(P) |\phi_A \phi_B \beta \rangle \,,
\label{ec:funonda}
\end{equation}
where $|\psi_\alpha\rangle$ are $b\bar{s}$ eigenstates of the two-body 
Hamiltonian, $\phi_{M}$ are wave functions associated with the $A$ and $B$ 
mesons, $|\phi_A \phi_B \beta \rangle$ is the two meson state with $\beta$ 
quantum numbers coupled to total $J^{P}$ quantum numbers and $\chi_\beta(P)$ is 
the relative wave function between the two mesons in the molecule. To derive 
the meson-meson interaction from the quark-antiquark interaction we use the 
Resonating Group Method (RGM)~\cite{Tang:1978zz}.

The coupling between the quark-antiquark and meson-meson sectors requires the
creation of a light quark pair. The operator associated with this process 
should describe also the open-flavour meson strong decays and is given 
by~\cite{Segovia:2012cd}
\begin{equation}
\begin{split}
T =& -\sqrt{3} \, \sum_{\mu,\nu}\int d^{3}\!p_{\mu}d^{3}\!p_{\nu}
\delta^{(3)}(\vec{p}_{\mu}+\vec{p}_{\nu})\frac{g_{s}}{2m_{\mu}}\sqrt{2^{5}\pi}
\,\times \\
&
\times \left[\mathcal{Y}_{1}\left(\frac{\vec{p}_{\mu}-\vec{p}_{\nu}}{2}
\right)\otimes\left(\frac{1}{2}\frac{1}{2}\right)1\right]_{0}a^{\dagger}_{\mu}
(\vec{p}_{\mu})b^{\dagger}_{\nu}(\vec{p}_{\nu}) \,,
\label{eq:Otransition2}
\end{split}
\end{equation}
where $\mu$ $(\nu)$ are the spin, flavour and colour quantum numbers of the
created quark (antiquark). The spin of the quark and antiquark is coupled to
one. The ${\cal Y}_{lm}(\vec{p}\,)=p^{l}Y_{lm}(\hat{p})$ is the solid harmonic
defined in function of the spherical harmonic. We fix the relation of $g_{s}$ 
with the dimensionless constant giving the strength of the quark-antiquark pair 
creation from the vacuum as $\gamma=g_{s}/2m$, being $m$ the mass of the 
created quark (antiquark).

From the operator in Eq.~(\ref{eq:Otransition2}), we define the transition 
potential $h_{\beta \alpha}(P)$ within the $^{3}P_{0}$ model 
as~\cite{Ortega:2012rs} 
\begin{equation}
\langle \phi_{A} \phi_{B} \beta | T | \psi_\alpha \rangle =
P \, h_{\beta \alpha}(P) \,\delta^{(3)}(\vec P_{\rm cm}) \,,
\label{Vab}
\end{equation}
where $P$ is the two-meson relative momentum.

The usual version of the $^{3}P_{0}$ model gives vertices that are too hard 
specially when we work at high momenta. Following the suggestion of 
Ref.~\cite{Morel:2002vk}, we use a momentum dependent form factor to truncate 
the vertex as
\begin{equation}
\label{Vab mod}
h_{\beta \alpha}(P)\to h_{\beta \alpha}(P)\times 
e^{-\frac{P^2}{2\Lambda^2}} \,,
\end{equation}
where $\Lambda=0.84\,{\rm GeV}$ is the value used herein.

Adding the coupling with bottom-strange states we end-up with the 
coupled-channels equations
\begin{equation}
\begin{split}
&
c_\alpha M_\alpha +  \sum_\beta \int h_{\alpha\beta}(P) \chi_\beta(P)P^2 dP = E
c_\alpha\,, \\
&
\sum_{\beta}\int H_{\beta'\beta}(P',P)\chi_{\beta}(P) P^2 dP + \\
&
\hspace{2.50cm} + \sum_\alpha h_{\beta'\alpha}(P') c_\alpha = E
\chi_{\beta'}(P')\,,
\label{ec:Ec-Res}
\end{split}
\end{equation}
where $M_\alpha$ are the masses of the bare $b\bar{s}$ mesons and 
$H_{\beta'\beta}$ is the RGM Hamiltonian for the two-meson states obtained from
the $q\bar{q}$ interaction. Solving the coupling with the $b\bar{s}$ states, we
arrive to a Schr\"odinger-type equation
\begin{equation}
\begin{split}
\sum_{\beta} \int \big( H_{\beta'\beta}(P',P) + &
V^{\rm eff}_{\beta'\beta}(P',P) \big) \times \\
&
\times \chi_{\beta}(P) {P}^2 dP = E \chi_{\beta'}(P') \,,
\label{ec:Ec1}
\end{split}
\end{equation}
where
\begin{equation}
V^{\rm eff}_{\beta'\beta}(P',P;E)=\sum_{\alpha}\frac{h_{\beta'\alpha}(P')
h_{\alpha\beta}(P)}{E-M_{\alpha}} \,.
\end {equation}


\begin{table}[!t]
\begin{center}
\begin{tabular}{ccccc}
\hline
\hline
\tstrut
State & $J^{P}$ & The. $(\alpha_{s})$ &  The. $(\alpha_{s}^{2})$ & Exp. \\
\hline
\tstrut
$B_{s}$               & $0^{-}$ & $5348$ & $5348$ & $5366.7\pm0.4$ \\
$B_{s}^{\ast}$        & $1^{-}$ & $5393$ & $5393$ & $5415.8\pm1.5$ \\
\hline
$B_{s0}^{\ast}$       & $0^{+}$ & $5851$ & $5801$ & - \\
$B_{s1}^{\prime}$     & $1^{+}$ & $5883$ & $5858$ & - \\
$B_{s1}(5830)$        & $1^{+}$ & $5841$ & $5850$ & $5828.40\pm0.41$ \\
$B_{s2}^{\ast}(5840)$ & $2^{+}$ & $5856$ & $5867$ & $5839.98\pm0.20$ \\
\hline
\hline
\end{tabular}
\caption{\label{tab:1loopBs} Masses, in MeV, of the low-lying $P$-wave 
bottom-strange mesons predicted by the constituent quark model $(\alpha_{s})$
and those including one-loop corrections to the one-gluon exchange potential 
$(\alpha_{s}^{2})$. For completeness, our predictions for the $0^{-}$ and 
$1^{-}$ states that belong to the $j_{q}^{P}=1/2^{-}$ doublet are included. 
Experimental data are taken from Ref.~\cite{Agashe:2014kda}.}
\end{center}
\end{table}

\section{Results}
\label{sec:Bs}

The heavy-quark doublet $j_{q}^{P} = 3/2^{+}$ is well established in the PDG, 
with the $B_{s1}(5830)$ and $B_{s2}^{\ast}(5840)$ mesons belonging to this 
doublet. Table~\ref{tab:1loopBs} shows the predicted masses within the naive 
quark model; one can see our results taking into account the one-gluon exchange 
potential $(\alpha_{s})$ and including its one-loop corrections 
$(\alpha_{s}^{2})$. In both cases our values are slightly higher than the 
experimental figures but compatible.

\begin{table}[!t]
\begin{center}
\begin{tabular}{cccccc}
\hline\hline
State & Mass & Width & ${\cal P}[q\bar{q}\,(^{3}P_{0})]$ & ${\cal 
P}[BK(S-wave)]$ \\
& (MeV) & (MeV) & (\%) & (\%) \\
\hline
$B_{s0}^{\ast}$ & $5741.4$ & $0$     & $61.2$ &  $38.8$ \\
$B_{s0}^R$       & $5877.9$ & $303.5$ & $83.7$ &  $16.3$ \\
\hline\hline
\end{tabular}
\caption{\label{tab:Bs0Pro} Masses, widths and probabilities of the different 
Fock components for the states found in the $J^P=0^{+}$ $b\bar{s}$ sector.}
\end{center}
\end{table}

The mass of the $B_{s0}^{\ast}$ state obtained using the naive quark model and 
without the one-loop spin corrections to the OGE potential is $5851\,{\rm 
MeV}$, which is compatible with the quark model predictions shown in 
Table~\ref{tab:Bs0Ene} but, nevertheless, is much higher than the average value 
predicted by lattice and EFT approaches. The mass associated to the 
$B_{s0}^{\ast}$ state is sensitive to the $\alpha_{s}^{2}$-corrections of the 
OGE potential. This effect brings down its mass about $50\,{\rm MeV}$ but 
still is incompatible with the lattice and EFT results. The mass-shift due to 
the $\alpha_{s}^{2}$-corrections allows the $0^{+}$ state to be closer to the 
$BK$ threshold. This makes the $BK$ coupling a relevant dynamical mechanism in 
the formation of the $B_{s0}^{\ast}$ bound state. When we couple the $0^{+}$ 
$b\bar{s}$ ground state with the $BK$ threshold the mass goes down towards 
$5741\,{\rm MeV}$ (Table~\ref{tab:Bs0Pro}), in good agreement with lattice and 
EFT estimations.

We turn now to discuss the probabilities of the different Fock components in 
the physical state. The lattice-regularised QCD study of 
Ref.~\cite{Lang:2015hza} is only able to remark that both quark-antiquark and 
meson-meson lattice interpolating fields have non-vanishing overlaps with the 
physical state. Our wave function probabilities are given in 
Table~\ref{tab:Bs0Pro} which reflects that the $B_{s0}^{\ast}$ meson is mostly 
of quark-antiquark nature. This is in agreement with the fact that 
lattice-regularised QCD computations observe this state even with only 
$q\bar{q}$ interpolators~\cite{Gregory:2010gm}.

In our model the probability of the $BK$ state depends basically on three 
quantities: the bare meson mass, the $^{3}P_{0}$ coupling constant and the 
residual $BK$ interaction. Obviously, as neither of the three are observables, 
they can take different values depending on the dynamics, making the results, 
and hence the $BK$ probability, model dependent. No model parameters have been 
changed from our study of the $D_{s0}^{\ast}(2317)$ meson in 
Ref.~\cite{Ortega:2016mms}. Moreover, the above mentioned quantities have been 
constrained in previous works by reproducing other physical observables like 
strong decays~\cite{Segovia:2012cd} (the $^3P_0$ coupling constant), 
bottomonium spectrum (the bare mass)~\cite{Segovia:2016xqb} and $NN$ and 
$p\bar p$ interactions (the $BK$ residual interaction)~\cite{Entem:2000mq, 
Entem:2006dt}.

The scattering length is sensitive to the $BK$ compositeness in the 
$B_{s0}^{\ast}$ wave function. Therefore, to complete the analysis of the 
$J^P=0^{+}$ $b\bar{s}$ sector, we have calculated the value of the scattering 
length from the value of the $T$-matrix at threshold, obtaining 
$a_{0}^{BK}=-1.18\,{\rm fm}$. This value is compatible with the one reported by 
Ref.~\cite{Lang:2015hza}, that is, $a_0^{BK}=(-0.85\pm0.10)\,{\rm fm}$, where 
the difference may originate from the binding energies predicted in each study.

In addition to the $B_{s0}^{\ast}$ state below the $BK$ threshold, we find a 
resonance with mass $5.88\,{\rm GeV}$ and width $300\,{\rm MeV}$ which is 
denoted as $B_{s0}^{R}$ in Table~\ref{tab:Bs0Pro}. The resonance is a 
$\sim\!84\,\%$ $b\bar s$ state which comes from the residual bare pole, that 
does not disappear but it is dressed with the $BK$ interaction and quickly 
moves into the complex plane. This resonance can be potentially observed in the 
$BK$ channel, but its large width is a handicap for the experiments.

\begin{table*}[!t]
\begin{center}
\begin{tabular}{ccccccc}
\hline
\hline
State & Mass & Width & ${\cal P}[q\bar{q}\,(^{1}P_{1})]$ & ${\cal 
P}[q\bar{q}\,(^{3}P_{1})]$ & ${\cal P}[B^{\ast}K(S-wave)]$ & ${\cal 
P}[B^{\ast}K(D-wave)]$ \\
& (MeV) & (MeV) & (\%) & (\%) & (\%) & (\%) \\
\hline
$B_{s1}^{\prime}$ & $5792.5$ & $0.000$ &  $13.5\%$ & $42.3\%$ & $44.2\%$ & - \\
$B_{s1}(5830)$ & $5850.0$ & $0.024$ & $37.2\%$ &  $12.8\%$ & $50.0\%$ & - \\
$B_{s1}^R$ & $5940.4$ & $271.3$ & $21.4\%$ & $58.8\%$ & $19.8\%$ & - \\
\hline
$B_{s1}^{\prime}$ & $5792.5$ & $0.000$ &  $13.2\%$ & $42.6\%$ & $44.2\%$ & 
$0.0\%$ \\
$B_{s1}(5830)$ & $5832.9$ & $0.058$ & $35.4\%$ &  $12.1\%$ & $15.9\%$ & 
$36.6\%$\\
$B_{s1}^R$ & $5940.4$ & $271.3$ & $21.4\%$ & $58.8\%$ & $19.8\%$ & $0.0\%$ \\
\hline
\hline
\end{tabular}
\caption{\label{tab:Bs1Pro} Masses, widths and probabilities of the different 
Fock components for the states found in the $J^P=1^{+}$ $b\bar{s}$ sector. 
Results with and without coupling of the $D$-wave $B^{\ast}K$ channel are 
listed.}
\end{center}
\end{table*}

The quark model including $\alpha_{s}^{2}$-corrections to the OGE potential 
predicts that the states corresponding to the $B_{s1}^{\prime}$ and 
$B_{s1}(5830)$ mesons are almost degenerated, with masses close to the 
experimentally observed mass of the $B_{s1}(5830)$. This $B_{s1}^{\prime}$ 
result goes in the same line than the ones predicted by other phenomenological 
models (see Tables~\ref{tab:Bs1Ene} and~\ref{tab:1loopBs}). However, the 
average of phenomenological model predictions, including our naive one, is 
around $80\,{\rm MeV}$ higher than lattice studies and EFT estimations.

We couple the two $1^{+}$ $b\bar{s}$ states associated with the 
$B_{s1}^{\prime}$ and $B_{s1}(5830)$ mesons with the $B^{\ast}K$ threshold. 
Following lattice criteria, we couple first the $B^{\ast}K$ channel in an 
$S$-wave. One can see in Table~\ref{tab:Bs1Pro} that the state associated with 
the $B_{s1}^{\prime}$ meson goes down in the spectrum and it is located below 
$B^{\ast}K$ threshold. Our value, $5793\,{\rm MeV}$, is compatible within 
errors with lattice data and with most of the EFT estimates. The state 
associated with the $B_{s1}(5830)$ meson is almost insensitive to the 
$B^{\ast}K$ $S$-wave coupling. The mass predicted in this case is 
$m_{B_{s1}(5830)}=5850\,{\rm MeV}$, which is the same than the one obtained 
using quark model without coupling. This is because the $B_{s1}(5830)$ state is 
the $J^{P}=1^{+}$ member of the $j_{q}^{P}=3/2^+$ doublet predicted by HQS and 
thus it couples mostly in a $D$-wave to the $B^{\ast}K$ threshold. It is 
important to highlight that the lattice computation of Ref.~\cite{Lang:2015hza} 
does not take into account the coupling of the $B_{s1}(5830)$ meson to the 
$B^{\ast}K$ threshold either in $S$- or $D$-wave. This is because the coupling 
of the $B^{\ast}K$ in $D$-wave is not easy to implement and, moreover, they 
assume that the coupling to $B^{\ast}K$ in $S$-wave is small.

The coupling in $D$-wave of the $B^{\ast}K$ threshold is trivially implemented 
in our approach. Table~\ref{tab:Bs1Pro} shows that the state associated with 
the $B_{s1}^{\prime}$ meson experience a negligible modification. This is 
understandable because it is almost the $|1/2,1^{+}\rangle$ eigenstate of HQS. 
The state associated with $B_{s1}(5830)$ meson suffers a moderate mass-shift 
with a final mass of $5833\,{\rm MeV}$. This value compares nicely with the one 
reported in Ref.~\cite{Lang:2015hza}: $(5831\pm10)\,{\rm MeV}$, but the 
coupling in $D$-wave of the $B^{\ast}K$ threshold in the lattice result needs 
to be incorporated in order to make a stronger statement. In any case, our 
result for the $B_{s1}(5830)$ is compatible with other theoretical approaches 
and with the experimental value reported in PDG. 

Table~\ref{tab:Bs1Pro} shows the probabilities of the different Fock components 
in the physical $B_{s1}^{\prime}$ and $B_{s1}(5830)$ states. When the 
$B^{\ast}K$ threshold is coupled in an $S$-wave, the meson-meson component is 
$44\%$ for the $B_{s1}^{\prime}$ and $50\%$ for the $B_{s1}(5830)$. It is also 
interesting to point out that the relation between the quark-antiquark partial 
wave components is close to the predictions of HQS, being the $B_{s1}^{\prime}$ 
meson a dominant $j_q=1/2$ state ($81\%$) and the $B_{s1}(5830)$ a $j_q=3/2$ 
state ($83\%$). One can also see in Table~\ref{tab:Bs1Pro} that the coupling of 
the $B^{\ast}K$ threshold in $D$-wave has very little effect for the formation 
of the $B_{s1}^{\prime}$ meson, keeping the probabilities pretty much the same. 
However, the $D$-wave coupling of the $B^{\ast}K$ channel is relatively 
important in the formation of the $B_{s1}(5830)$ with a contribution to its 
physical wave function of about $37\%$. Moreover, in the case of the 
$B_{s1}(5830)$, the distribution of the meson-meson component in $S$- and 
$D$-wave channels does not affect very much the probabilities of the 
quark-antiquark components, being still dominant the $j_q=3/2$ HQS state.

As in the $J^{P}=0^{+}$ $b\bar{s}$ sector, we can predict the value of the 
scattering length for the $S$-wave $B^{\ast}K$ channel, $a_0^{B^\ast 
K}=-1.35\,{\rm fm}$, and find that is slightly higher but compatible with the 
$a_0^{B^{\ast}K}=(-0.97\pm0.16)\,{\rm fm}$ reported by Ref.~\cite{Lang:2015hza}. 
Again, the different mass for the $B_{s1}^{\prime}$ predicted in each study can 
account for the difference in the predicted scattering lengths.

In the $J^{P}=1^{+}$ $b\bar{s}$ sector we also find a broad resonance above 
the $B^{\ast}K$ threshold. It is originated from the $j_q^P=1/2^+$ component of 
the bare $b\bar s$ pole, which is largely dressed with the $B^\ast K$ $S$-wave 
interaction. This resonance, labelled $B_{s1}^R$ in Table~\ref{tab:Bs1Pro}, is 
located at the mass region of $5.94\,{\rm GeV}$ and has a width of $271\,{\rm 
MeV}$, which complicates but not forbids its possible 
experimental observation.

Only quark-antiquark operators were used in the lattice 
study~\cite{Lang:2015hza} of the $B_{s2}^{\ast}(5840)$ meson. They obtained a 
value of $(5853\pm13)\,{\rm MeV}$ for the mass. This is in qualitative agreement 
with experiment and with our naive quark model prediction, confirming that this 
state can be described well within the $b\bar{s}$ picture.

Finally, let us clarify that we have omitted the possible coupling of the 
states to the $B_{s}^{(\ast)}\pi$ channels since they violate isospin and thus 
the effect should be subleading. We also neglect effects coming from the 
$B_{s}^{(\ast)}\eta$ coupling, partially motivated by the threshold lying ${\cal 
O}(140\,{\rm MeV})$ above the $B^{(\ast)}K$ threshold. Similar criteria has 
been followed by the lattice analysis of Ref.~\cite{Lang:2015hza}.


\section{EPILOGUE}
\label{sec:epilogue}

Within the formalism of a nonrelativistic constituent quark model in which 
quark-antiquark and meson-meson components are incorporated, we have performed 
a coupled-channel computation taking into account the $J^{P} = 0^{+}$, $1^{+}$ 
and $2^{+}$ bottom-strange states predicted by the naive quark model and the 
$BK$ and $B^{\ast}K$ thresholds. Our method allows to introduce the coupling 
with the $D$-wave $B^{\ast}K$ channel and to compute the probabilities 
associated with the different Fock components of the physical state, features  
which cannot be addressed nowadays by other theoretical approaches.

Our study has been motivated by the fact that there are no experimental 
evidences of the $b\bar s$ mesons which belong to the doublet 
$j_{q}^{P}=1/2^{+}$ and, as it has been seen in the $c\bar s$ 
sector~\cite{Ortega:2016mms}, the naive theoretical result can be wrong by more 
than $100\,{\rm MeV}$. In order to keep the predictive power of the formalism we 
do not change any parameter of the calculation in Ref.~\cite{Ortega:2016mms}. 
Moreover, it is worth to emphasize again that the quark model has been applied 
to a wide range of hadronic observables and thus the model parameters are 
completely constrained.

The level assigned to the $B_{s0}^{\ast}$ meson within the naive 
quark model is much higher than the ones predicted by lattice and EFT 
approaches. However, it has been shown that the value is compatible with other 
phenomenological model predictions. The one-loop corrections to the OGE 
potential brings down this level and locates it slightly above the $BK$ 
threshold. This makes the coupling with the nearby threshold to acquire an 
important dynamical role. When coupling, the level is down-shifted again 
towards the average mass obtained by lattice and EFT formalisms. We predict 
a probability of around $40\%$ for the $BK$ component of the $B_{s0}^{\ast}$ 
wave function. Lattice QCD can only state that both quark-antiquark and 
meson-meson operators have important overlaps with the physical state.

The $B_{s1}^{\prime}$ and $B_{s1}(5830)$ mesons appear almost degenerated using 
the naive quark model that includes the one-loop corrections to the OGE 
potential. We have coupled the two $1^{+}$ $b\bar{s}$ states associated with 
the $B_{s1}^{\prime}$ and $B_{s1}(5830)$ mesons with the $B^{\ast}K$ threshold. 
When coupling the $B^{\ast}K$ channel in a $S$-wave, the $B_{s1}^{\prime}$ 
state goes down in the spectrum and it is located below $B^{\ast}K$ threshold 
with a mass compatible with lattice and EFT predictions. The $B_{s1}(5830)$ 
meson is almost insensitive to this coupling because it is the 
$|3/2,1^{+}\rangle$ state predicted by HQS and thus couples mostly in a $D$-wave 
to the $B^{\ast}K$. When such coupling is included the state associated 
with $B_{s1}(5830)$ meson suffers a moderate mass-shift and it is in very good 
agreement with other theoretical approaches and with the value reported in PDG. 
We observe that the meson-meson component is around $50\%$ for both 
$B_{s1}^{\prime}$ and $B_{s1}(5830)$ mesons, taking the quark-antiquark
partial waves the other $50\%$.

It is worth mentioning that, while finishing the preparation of this work, 
a new study was published~\cite{Albaladejo:2016lbb} supporting the idea that 
the $B_{s0}^\ast$ and the $B_{s1}^{\prime}$ should be in the mass region 
of $5720$ and $5770$ MeV, respectively. Such values are compatible with 
other EFT predictions and with the results presented in this work.

Finally, the mass of the $B_{s2}^{\ast}(5840)$ meson is predicted reasonably 
well within our quark model approach taking into account only quark-antiquark 
degrees-of-freedom. The same conclusion has been drawn by lattice-regularised 
QCD computations.




\begin{acknowledgments}
This work has been partially funded by Ministerio de Ciencia y Tecnolog\'\i a 
under Contract no. FPA2013-47443-C2-2-P, by the Spanish Excellence Network 
on Hadronic Physics FIS2014-57026-REDT, and by the Junta de Castilla y Le\'on 
under Contract no. SA041U16. P.G.O. acknowledges the financial support from the 
Spanish Ministerio de Econom\'ia y Competitividad and European FEDER funds 
under the contract no. FIS2014-51948-C2-1-P. J.S. acknowledges the financial 
support from Alexander von Humboldt Foundation.
\end{acknowledgments}


\bibliography{MolecularComponents_Bs0Bs1}

\end{document}